\documentclass[prx,aps,twocolumn,amssymb,superscriptaddress,showpacs,floatfix]{revtex4-1}
\usepackage{amssymb}

\usepackage[pdftex]{graphicx}
\usepackage{epstopdf}

\begin{document}

\title{Short-Range Correlations in Magnetite above the Verwey Temperature}

\author{Alexey Bosak}
\affiliation{European Synchrotron Radiation Facility, 6 rue Jules Horowitz,
             F-38043 Grenoble Cedex, France}
             
\author{Dmitry Chernyshov}
\affiliation{Swiss-Norwegian Beam Lines at ESRF, 6 rue Jules Horowitz,
             F-38043 Grenoble Cedex, France}             

\author{Moritz Hoesch}
\affiliation{Diamond Light Source, Harwell Campus, Didcot OX11 0DE, Oxfordshire, England}

\author{Przemys\l{}aw Piekarz}
\affiliation{Institute of Nuclear  Physics, Polish Academy of Sciences,
             Radzikowskiego 152, PL-31342 Krak\'ow, Poland }

\author{Mathieu Le Tacon}
\affiliation{Max-Planck-Institut f\"ur Festk\"orperforschung,
             Heisenbergstrasse 1, D-70569 Stuttgart, Germany}

\author{Michael~Krisch}
\affiliation{European Synchrotron Radiation Facility, 6 rue Jules Horowitz,
             F-38043 Grenoble Cedex, France}

\author{Andrzej Koz\l{}owski}
\affiliation{Faculty of Physics and Applied Computer Science,
             AGH-University of Science and Technology, \\
             Aleja Mickiewicza 30, PL-30059 Krak\'ow, Poland}

\author{Andrzej M. Ole\'{s}}
\affiliation{Max-Planck-Institut f\"ur Festk\"orperforschung,
             Heisenbergstrasse 1, D-70569 Stuttgart, Germany}
\affiliation{Marian Smoluchowski Institute of Physics, Jagellonian University,
             Reymonta 4, PL-30059 Krak\'ow, Poland}

\author{Krzysztof Parlinski}
\affiliation{Institute of Nuclear Physics, Polish Academy of Sciences,
             Radzikowskiego 152, PL-31342 Krak\'ow, Poland }

\date{\today}

\begin{abstract}
Magnetite, Fe$_3$O$_4$, is the first magnetic material discovered and 
utilized by mankind in Ancient Greece, yet it still attracts attention 
due to its puzzling properties. This is largely due to the quest for a 
full and coherent understanding of the Verwey transition that occurs at 
$T_V=124$ K and is associated with a drop of electric conductivity and 
a complex structural phase transition. A recent detailed analysis of 
the structure, based on single crystal diffraction, suggests that 
the electron localization pattern contains linear three-Fe-site units, 
the so-called trimerons. Here we show that whatever the electron 
localization pattern is, it partially survives up to room temperature 
as short-range correlations in the high-temperature cubic phase, 
easily discernible by diffuse scattering. Additionally, {\it ab initio} 
electronic structure calculations reveal that characteristic features 
in these diffuse scattering patterns can be correlated with the Fermi 
surface topology.
\end{abstract}


\maketitle

\section{Introduction}

Discovered in the first half of the 20th century, the Verwey transition 
in magnetite \cite{verwey} remains one of the most intriguing phenomena 
in solid state physics. Magnetite is a ferrimagnetic spinel with 
anomalously high Curie temperature $T_C=850$ K. Hence, it is viewed 
as an ideal candidate for room temperature spintronic applications. It 
crystalizes in the inverse spinel cubic structure, with two types of Fe 
sites: the tetrahedral $A$ sites and the octahedral $B$ ones 
\cite{iizumi,imada,walz1,walz2,att1,att2,garcia}. 
At $T_V = 124$ K, a first order phase transition occurs as the 
electric conductivity drops by two orders of magnitude \cite{verwey} 
with the simultaneous change of the crystal structure from the cubic to 
monoclinic symmetry \cite{iizumi} and with spectacular anomalies in 
practically all physical characteristics 
\cite{imada,walz1,walz2,att1,att2,garcia}. The low-temperature 
structure of magnetite, as deduced from recent studies, is identified 
to be of monoclinic $Cc$ space group symmetry, with complex 
displacement pattern \cite{senn}. 

In recent years, the main research effort was focused on the 
low-temperature phase in order to elucidate the character of charge 
ordering (CO) first proposed by Verwey as the primary mechanism of the 
phase transition. From these investigations a complex picture of a 
low-symmetry state arises involving charge, orbital, and lattice 
degrees of freedom \cite{att1,att2,garcia,senn,leonov,jeng,prl,prb,pinto,Naz06,lorenzo,Sch08,blasco,tanaka,subias1,senn2}. 
Diffraction studies performed below $T_V$ have revealed a fractional 
charge disproportionation \cite{att1,att2}, with a CO and associated 
orbital ordering on $B$ sites, which can be explained by strong Coulomb 
interactions \cite{leonov,jeng} which amplify the coupling between 
$3d$ electrons and lattice deformation \cite{prl,prb,pinto}. This 
charge and orbital order was recently suggested to exist in the form of 
so-called trimerons, distributed over three octahedral Fe sites, 
and coupled to the lattice distortion \cite{senn}.

Above the Verwey transition, where the inverse spinel structure, 
with Fe$^{3+}$ ions in $A$ sites and $B$ sites in a mixed-valence 
Fe$^{2.5+}$ state is realized, several observations indicate the 
existence of short-range order of polaronic character
\cite{ihle,park,schrupp,subias2}. This correlated state is reflected in 
the critical softening, on cooling, of the $c_{44}$ elastic constant 
\cite{moran,schwenk}, softening of the surface phonons \cite{seikh}, 
critical diffuse scattering 
\cite{fujii,chiba,shapiro,yamada,aragon,siratori} and anomalous phonon 
broadening \cite{hoesch}. Furthermore, the quasi-elastic character of 
neutron scattering suggests low-energy fluctuations of the lattice 
distortions coupled to electrons \cite{shapiro,yamada}. 

This short-range order above the Verwey transition, its subtleties and 
its connection to the low-temperature phase is still not completely 
understood. Thus, to gain further insights into the transition, the 
observation of this short range order and its coupling to the 
electronic properties of the $t_{2g}$ minority-spin states of the Fe 
octahedral ($B$) atoms (which contribute to the metallic state above 
the Verwey transition) is crucial. We will show below that Fermi surface 
nesting features may be responsible for the observed short range order.

Typical spot-like diffuse scattering was reported a few Kelvin above 
$T_V$ at positions $q=(h,0,l+1/2)$, which become Bragg reflections 
below the phase transition \cite{fujii}. Another type of diffuse 
scattering with a disc-like shape was revealed close to the $\Gamma$ 
and $X$ points over a wide range of temperatures 
\cite{chiba,shapiro,yamada}. 
In order to clarify the exact shape and behavior of diffuse scattering, 
as well as to unambiguously demonstrate its relation to the low-symmetry 
structure, we have studied the evolution of x-ray diffuse scattering in 
magnetite as a function of temperature down to the Verwey transition and 
below. Thanks to the use of a state-of-the-art large area detector 
(PILATUS 6M), the detailed three-dimensional (3D) reciprocal space 
mapping could be performed revealing an extremely rich diffuse 
scattering pattern, inherited from the complex low-temperature structure 
below the Verwey transition.

The remaining of the paper is organized as follows. The experimental 
part is described in Sec. \ref{sec:exp}. In Sec. \ref{sec:gga} we give 
the details of the performed {\it ab initio} electronic structure 
calculations which serve to determine the Fermi surface in the metallic 
state. In Sec. \ref{sec:res} we present and discuss the obtained 
results of diffuse scattering experiments (Sec. \ref{sec:dif}), and
present the consequences of charge ordering (Sec. \ref{sec:co}). The 
main results for the Fermi surface of magnetite are given in Sec. 
\ref{sec:fs}. There we also report a remarkable agreement between the 
experimentally obtained nesting signatures in the reciprocal space of 
magnetite, and the calculated Fermi surface nesting. The paper is 
summarized in Sec. \ref{sec:summa}.

\section{Experiment}
\label{sec:exp}

The single crystalline magnetite samples were grown at Purdue University 
by the skull melter, crucibleless technique \cite{aragon2}. This allowed 
controlling of the oxygen partial pressure during growth, thereby 
ensuring that the melt remains within the stability range of the 
material. After preparation, the crystals were subjected to subsolidus 
annealing under CO/CO$_2$ gas mixtures to establish the appropriate 
metal/oxygen ratio \cite{aragon3}. Due to rapid quenching from high 
temperatures, this procedure generates octahedral defects and introduces 
stress \cite{walz3}. However, the stoichiometry is maintained and most 
of the low temperature electronic processes are not affected, as is 
evidenced by the sharp Verwey transition with high $T_V\sim124$ K, see
Figure 1 in the Supplemental Material \cite{suppl}.

Similar low-temperature (low-$T$) processes were revealed by microscopic 
probes \cite{chlan,novak} for samples prepared in a different way. 
Thus, electronic processes that cause diffuse scattering result from 
the intrinsic properties of magnetite, and not from particular 
preparation conditions.

Prior to the experiment, the nearly stoichiometric crystals 
(nominal $\delta = 0.00003$ and $\delta = -0.0001$ for 
Fe$_{3(1-\delta)}$O$_4$) were mechanically put to the shape of a needle 
and etched down to ~50 $\mu$m diameter with HCl in order to remove 
the damaged surface layer. Both samples gave identical results; the
data for $\delta= -0.0001$ are presented and referred to. 
It is noteworthy, that the diffuse patterns are stable with respect to 
the nonstoichiometry above $T_V$, while below the same pattern can 
coexist with the superstructure, see Figure 2 in the Supplemental 
Material \cite{suppl}.

The scattering data were collected at a room temperature, $T=T_V+2.5$ K  
and $T=T_V-2.5$ K, and a number of intermediate temperatures (155 K, 
195 K, 245 K) in shutter-less mode with the PILATUS 6M detector 
\cite{bronnimann}. The preliminary measurements were performed at 
beamline X06SA at the SLS, the follow-up data recording took place at 
beamline ID29 at the ESRF; in both cases a wavelength of 0.7 \AA\ was
employed. The crystal was mounted on a horizontal rotation stage, and 
the diffuse scattering patterns were recorded with an increment of 
$0.1^{\circ}$ over an angular range of 360$^{\circ}$ with   0.25 s
exposure per frame, i.e., 15 min per full dataset. 

The experimental geometry was refined with the help of the CrysAlis 
software \cite{oxford} that was also used for the preliminary data 
evaluation. The reconstruction of the selected reciprocal space layers 
was performed with a locally developed software. The reconstructed 
volume was averaged with its symmetrically equivalent orientations
employing the Laue symmetry of the average structure, thus improving 
the signal-to-noise ratio and removing the gaps between individual 
detector elements. The low-temperature data were collected utilizing 
an attenuator in order to avoid intensity saturation effects of the 
superstructure reflections.

\section{Electronic structure calculations}
\label{sec:gga}

Ab initio calculations of the Fermi surface (FS) were performed in the 
cubic inverse-spinel unit cell within the generalized-gradient 
approximation of the density functional theory (GGA/DFT) using the 
all-electron WIEN2k code \cite{blaha}. The linearized augmented plane 
wave basis with local orbitals was expanded to $k_{max}$ given by 
$r\cdot k_{max}=7$ outside the atomic spheres with radius $r=1.63$ a.u. 
for oxygen and $r=1.83$ a.u. for iron. 

A symmetry reduced grid of $21\times21\times21$ points in momentum 
space was used for convergence of the total energy. The lattice 
parameter was relaxed to 15.871 a.u. which agrees very well with the 
experimental value 15.862 a.u. A well-converged ferrimagnetic 
arrangement was obtained, with opposite orientations of magnetic 
moments in the $A$ and $B$ sites, as observed in magnetite below the 
Curie temperature $T_C=850$ K.

\section{Results and Discussion}
\label{sec:res}

\subsection{Diffuse scattering experiments}
\label{sec:dif}

\begin{figure}[t!]
\centerline{\includegraphics[width=8.2cm]{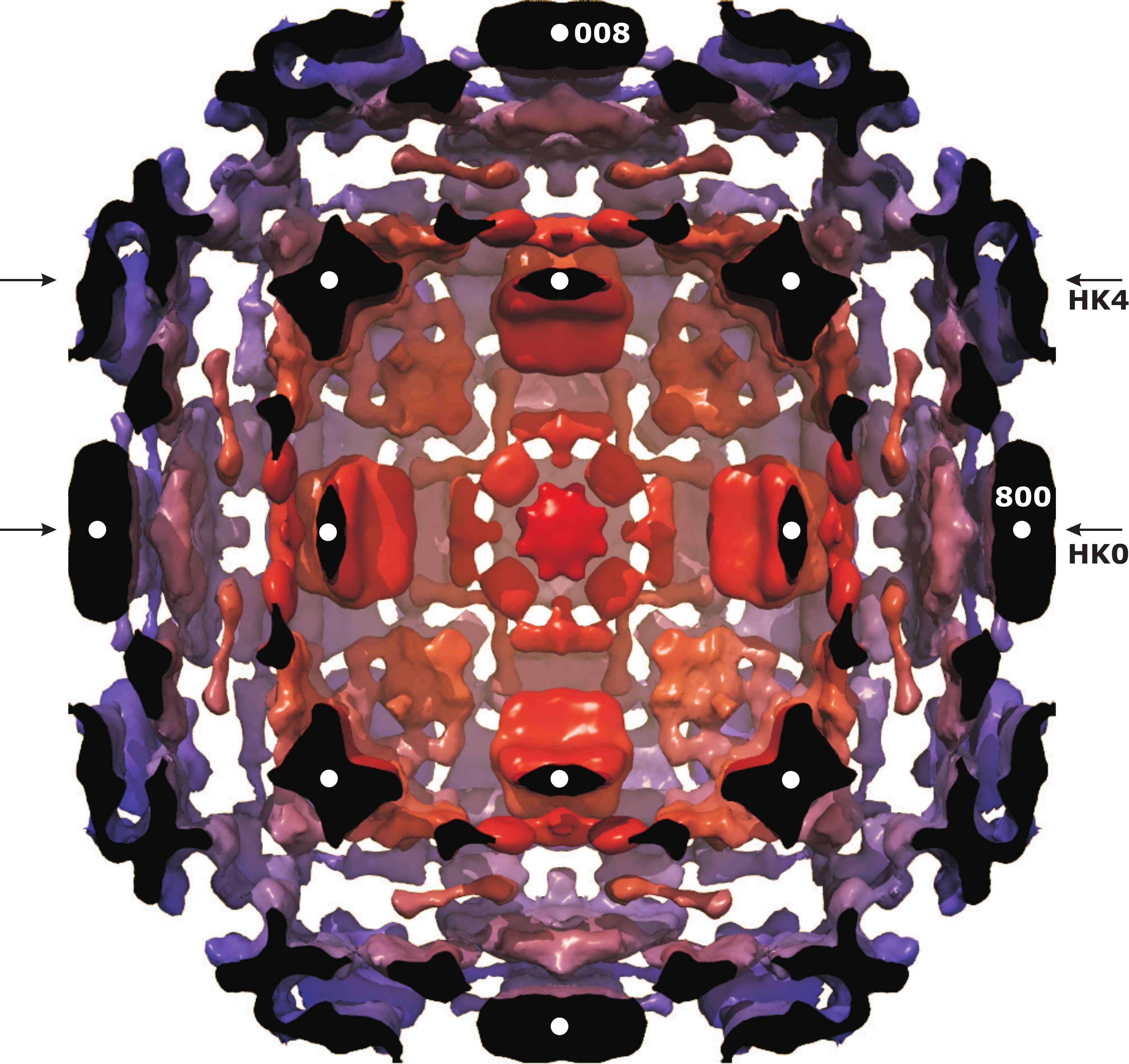}}
\caption{Isosurface representation of diffuse scattering in magnetite 
slightly above $T_V$. Color represents the distance to the (000) node; 
diffuse clouds in the proximity of weak Bragg spots are removed. 
The half-space above the H0L plane is removed. $|Q|$-dependent 
intensity scaling is applied for the purpose of better visualisation.
White circles mark strong Bragg reflections in the HOL plan; arrows 
denote HKO and HK4 cuts perpendicular to the image plane.}
\label{Fig1}
\end{figure}

Figure \ref{Fig1} shows the representative diffuse features of 
magnetite slightly above $T_V$ in form of isosurfaces. The strong main 
lattice Bragg spots are not visible here because their intensity is 
much higher than the value of the isosurfaces. The isolated diffuse 
clouds centered on weak Bragg reflections are removed, thus only keeping 
extended and/or interconnected fragments. Already this overview shows 
that the diffuse scattering cannot be reduced to simple objects, such 
as spots and discs, but in contrast, reveals a rich structure which is 
discussed in detail below.

\begin{figure}[t!]
\centerline{\includegraphics[width = .48\textwidth]{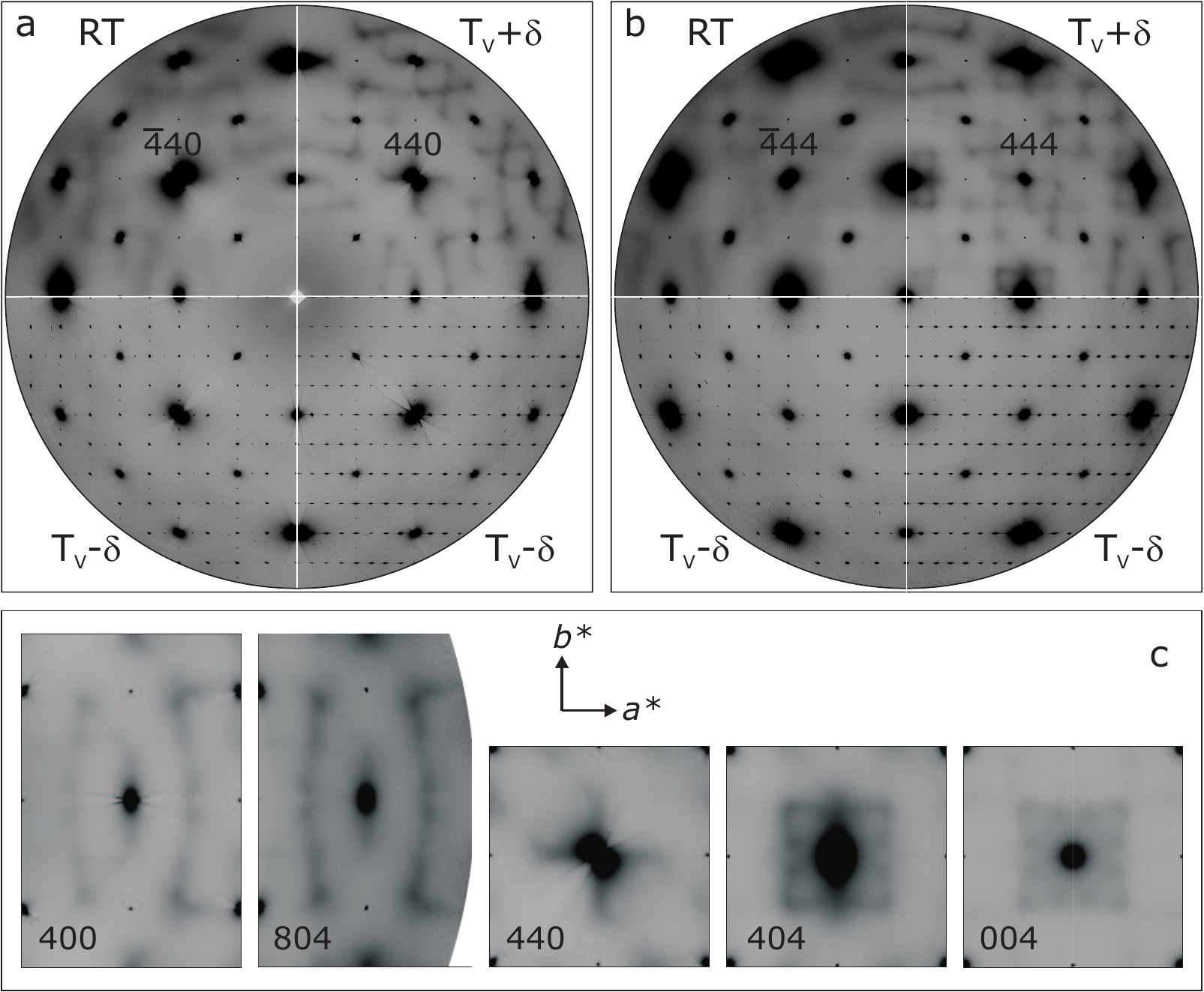}}
\caption{Diffuse scattering in magnetite at variable temperature. 
Magnetite reciprocal space
cuts HK0 (a) and HK4 (b) are shown at room temperature, slightly above 
($T_V+\delta$) and below ($T_V-\delta$) the Verwey transition, with 
$\delta\sim2.5$ K. Left bottom subpanels: sections perpendicular to the
cell doubling direction; right bottom subpanels: sections parallel to 
the cell doubling direction. Selected regions of interest (see text) 
are shown in (c) together with the unit cell vectors a* and b*.
The room temperature cut is scaled taking into account the thermal 
expansion for visualisation purposes; the labels in (c) denote the 
central Bragg spot. Cubic m3m (above $T_V$) and tetragonal
4/mmm (below $T_V$) Laue symmetries are applied.}
\label{Fig2}
\end{figure}

The results of the diffuse scattering measurements in magnetite for two 
representative reciprocal space cuts of the HK0 plane (left panel) and 
HK4 plane (right panel) are presented in Fig. \ref{Fig2}. Besides the 
contributions from thermal diffuse scattering (TDS), which arises 
predominantly from acoustic phonons and is centered around Bragg 
reflections, we observe other distinct diffuse features already at room 
temperature. On cooling, these features gradually become stronger 
and sharper (see Fig. \ref{Fig2}, upper panels). 

\begin{figure}[t!]
\centerline{\includegraphics[width = .48\textwidth]{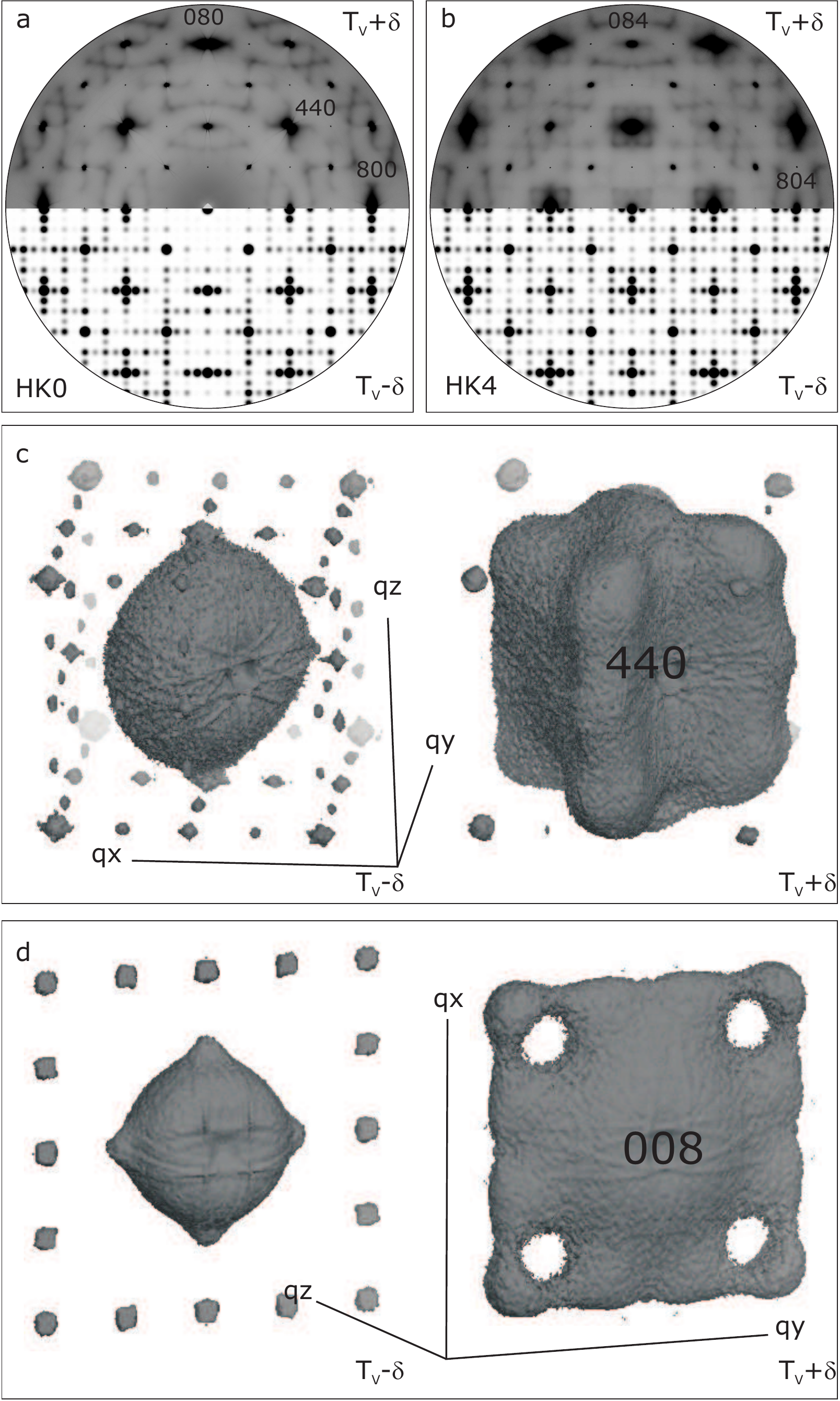}}
\caption{
Top panels --- Similarity of scattered intensity distribution in 
magnetite above and below the Verwey transition. Magnetite reciprocal 
space cuts, (a) HK0 and (b) HK4, are compared for temperatures above 
$T_V$ ($T_V+\delta$, upper subpanels) and below $T_V$ ($T_V-\delta$, 
lower subpanels), with $\delta\sim2.5$ K. Gaussian blurring is applied 
to the superstructure reflections; main reflections are replaced by 
constant intensity peaks to avoid masking the superstructure-related 
features. Middle and bottom panel ---
Isosurface representation of diffuse scattering in magnetite as given 
in the proximity of the (c) (440) and (d) (008) node. Indices and 
vector lengths are given in r.l.u. of the cubic structure. Cubic $m3m$ 
Laue symmetry is applied. Artificial twinning of the low-temperature 
data recovers the cubic symmetry.
}
\label{Fig3}
\end{figure}

Detailed mapping allows recovering the actual shape of these features, 
previously reported as spot-like and disk-like objects 
\cite{chiba,shapiro,yamada}. Among the most remarkable features, we can 
list “squares” centered on the strongest Bragg reflections of the spinel 
(400, 800, 440 and 448) structure and “arcs” (nearly symmetric pairs are 
visible around 400 and 804 reflections). Some distinct objects are shown 
in Fig. \ref{Fig2}(c). It is worth noting that the local maxima of 
diffuse intensity never appear at $X$ points, but are rather shifted 
aside, to incommensurate positions. 

Below the Verwey temperature the diffuse intensity collapses into the 
superstructure Bragg reflections. While the real symmetry of the 
low-temperature phase is monoclinic, the diffraction pattern appears 
similar to that of a tetragonal structure. The modulation vector 
corresponding to the doubling of the unit cell in the c direction 
appears in only one direction of the three $<$100$>$-type equivalents; 
thus, not more than 8 out of 24 twins allowed by the symmetry are 
apparent. This is illustrated by the lower panels of Fig. \ref{Fig2}, 
where the structure is considered to be pseudotetragonal with 
corresponding Laue symmetry operations applied.

\begin{figure}[t!]
\centerline{\includegraphics[width = .48\textwidth]{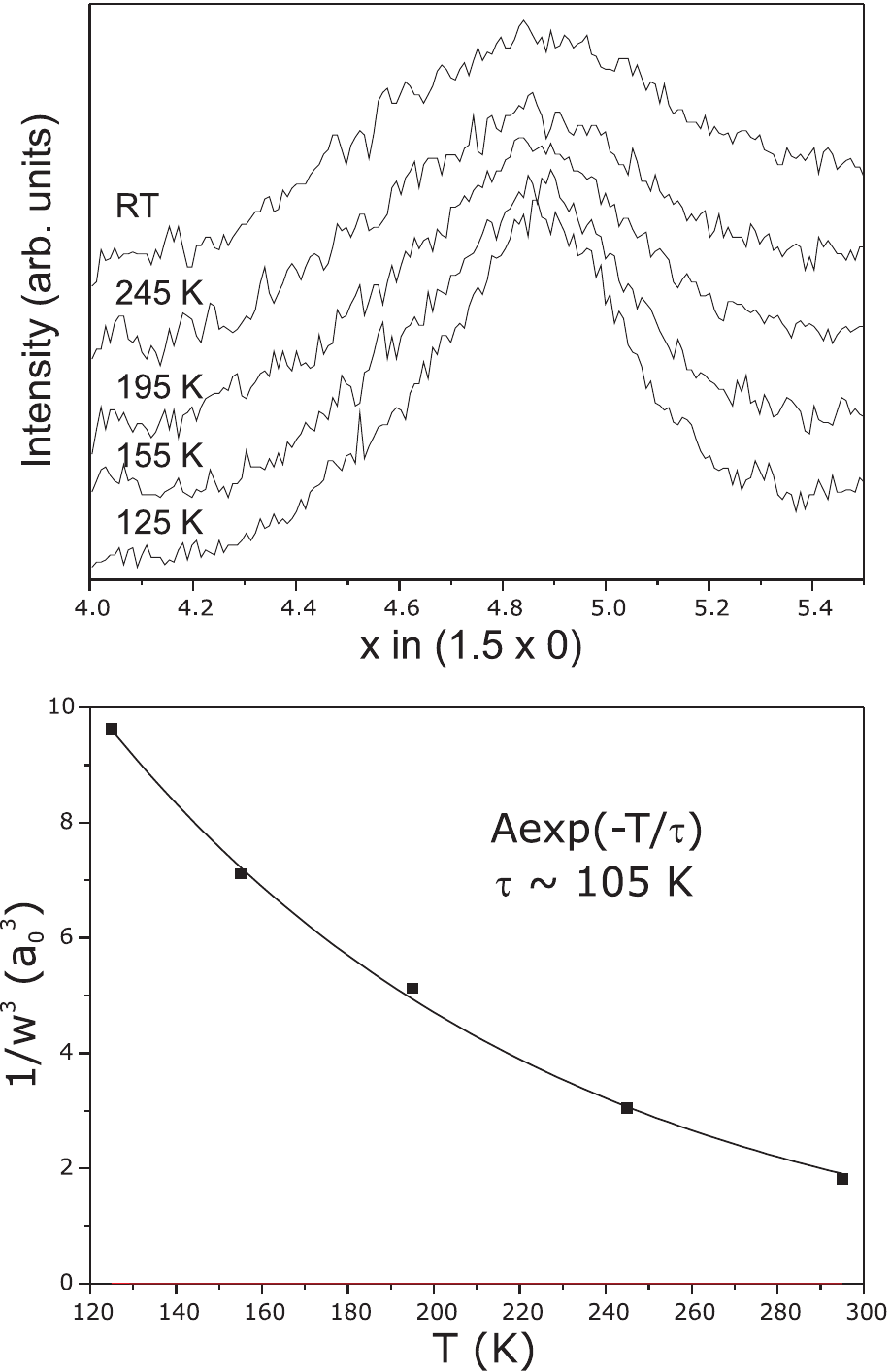}}
\caption{Temperature dependence of characteristic diffuse intensity 
in magnetite: 
(a) linear scans across the arc-like diffuse feature of 
magnetite for temperatures between $T_V$ and a room temperature; 
(b) the temperature decay of the derived $1/w^3$ value (symbols),
together with its exponential fit $A\exp(-T/\tau)$ (solid line).
Here $w$ is the width of the linear diffuse scattering scans,
and $a_0$ denotes the lattice parameter of the cubic lattice.
}
\label{Fig4}
\end{figure}

\subsection{Consequences of charge ordering}
\label{sec:co}

We will now compare the measured diffuse intensities with the 
scattering pattern of the ordered phase below $T_V$ (intensities 
taken from Ref. \cite{senn}). In the same linear greyscale, the 
superstructure reflections would be highly saturated. Thus, for 
the graphical representation we performed the following data 
transformations: 
(i) symmetrization by the operations of the $m3m$ point group, 
(ii) attenuation of cubic spinel Bragg reflections to a constant level, 
and 
(iii) convolution of all the reflections with a Gaussian profile. 
In this way the intensity of superstructure reflections is visualized 
via both the size and the intensity of spots; the result of this 
procedure is shown in Figs. \ref{Fig3}(a) and \ref{Fig3}(b) 
(lower panels) and compared to the diffuse dataset taken slightly above 
$T_V$. Comparision of the 3D intensity distribution above and below 
the transition is shown in Figs. \ref{Fig3}{c) and \ref{Fig3}(d).

The qualitative similarity is apparent: all diffuse features have their 
counterparts with proportional intensity in the low-$T$ diffraction at 
roughly the same $Q$, and vice versa. Subsets of strong reflections 
form the squares around the strongest spinel reflections as does the 
diffuse scattering above $T_V$; the diffuse “arcs”, in turn, transform 
to the chains of spots. This means that the basic short-range ordering 
pattern is inherited from the low-$T$ structure, while the transition 
is accompanied by the loss of commensurability (note that the $X$ 
points are avoided). 

The characteristic length estimated from the width of diffuse features 
varies from $\sim 2$ unit cells (u.c.) of the prototype cubic structure 
at $T_V+2.5$ K to a value slightly larger than $\sim 1$ u.c. at 
room temperature (see Fig. \ref{Fig4}). Thus, the ordering pattern 
cannot be reduced to the “trimeron” features \cite{senn}, but rather 
to complexes of trimerons. Therefore our study supports the polaron 
picture, and we can state that its structure is in reality much more 
complex than ever expected previously. 

\subsection{Fermi surface}
\label{sec:fs}

Cooling down to the Verwey temperature provokes condensation of these 
dynamic objects to the monoclinic structure with static charge ordering. 
The entropy change at $T_V$ should be further reduced compared to 
the trimeron-based estimates \cite{senn}. The local symmetry of the
complexes can be lower than cubic, but the average cubic symmetry is 
recovered by “nanotwinning”.

\begin{figure}[t!]
\centerline{\includegraphics[width = .48\textwidth]{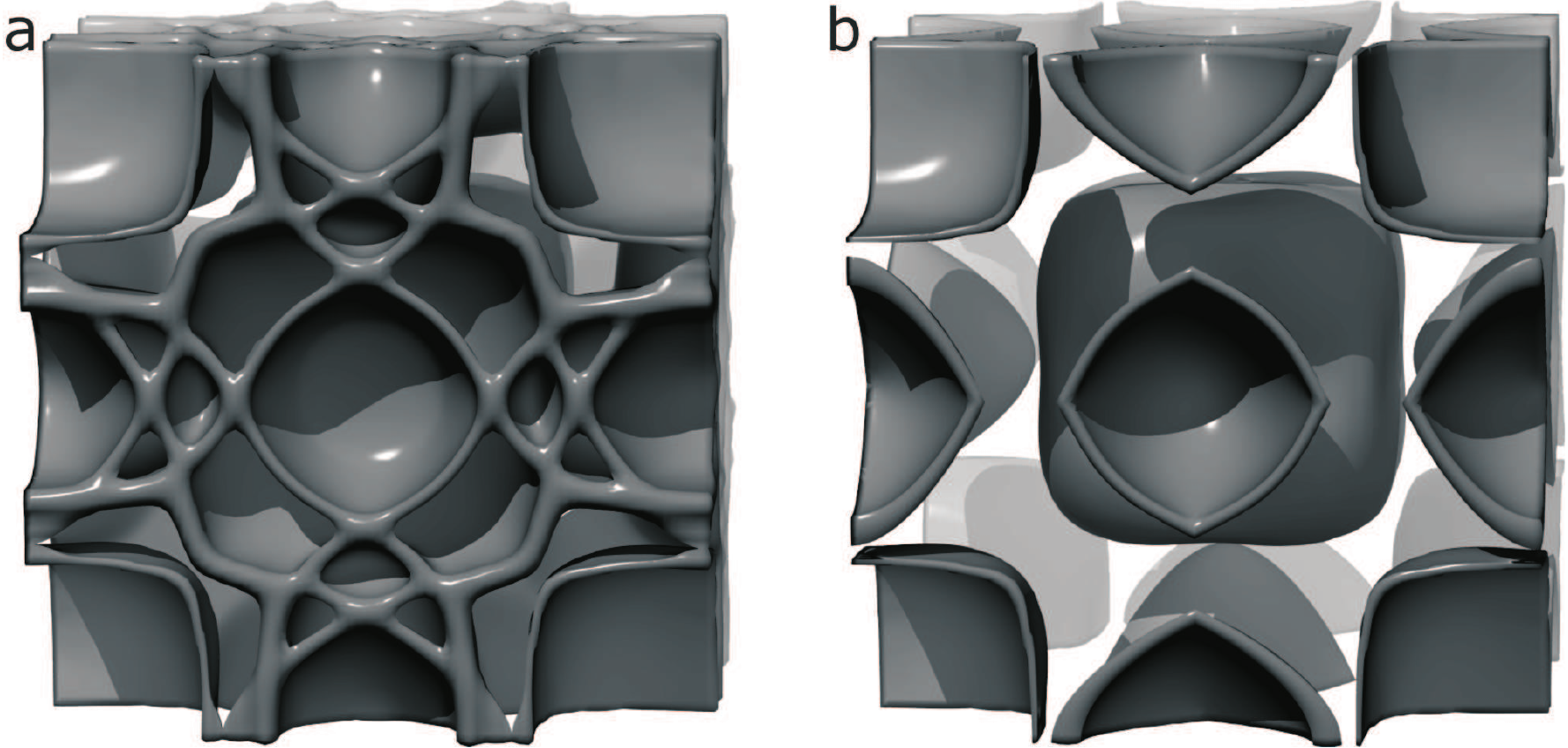}}
\caption{Fermi surface of magnetite in the cubic phase: 
(a) Total (multiband) Fermi surface of magnetite as derived from the 
spin-polarized electronic band structure using the WIEN2K code; 
(b) isolated minority-spin $t_{2g}$ band of interest.
}
\label{Fig5}
\end{figure}

Such a complex CO, which cannot be reduced to a few frozen phonon modes, 
can, at least partially, be explained by the Fermi surface topology. In 
fact, it has been shown that the short range order part of the diffuse 
scattering can reveal details about the electronic structure 
\cite{krivoglaz,reichert}. This is caused by an anomaly in the static 
susceptibility of the conduction electrons at scattering vectors, 
$k=2k_F$ ($k_F$ denotes the Fermi wave vector) and, therefore, in the 
Fourier transform of the pair-interaction potential. The effect can 
become particularly pronounced if different portions of the Fermi 
surface are connected by a single scattering vector 
(Fermi surface nesting). 

In order to validate our hypothesis, a number of nesting constructions 
based on our ab initio calculations were evaluated for intra- and 
interband transitions. For the nesting construction we used 
$|\nabla\chi_q|$, where
\begin{equation}
\chi_q=\sum_k\frac{n_F(\varepsilon_k)-n_F(\varepsilon_{k+q})}
{\varepsilon_k-\varepsilon_{k+q}}
\end{equation}
is the real part of the bare spin susceptibility (Lindhard function) 
\cite{ashcroft} at frequency $\omega\rightarrow0$; here
\begin{equation}
n_F(\varepsilon)=\frac{1}{\exp\left(\frac{\varepsilon}{k_BT}\right)+1} 
\end{equation}
stands for the Fermi-Dirac distribution function ($k_B$ is the 
Boltzmann constant), and $\varepsilon$ denotes the kinetic energy of 
the electrons (the chemical potential is set to zero). The total Fermi
surface has a quite complex shape, traced in Fig. \ref{Fig5}(a) as the 
isosurface of $\sum_n\exp(-\varepsilon_{k,n}^2/\alpha)$, where $n$ 
refers to the band number, and $\alpha$ is chosen to smear out the 
discretization artifacts. We have found that the scattering within the 
minority-spin $t_{2g}$ band shown in Fig. \ref{Fig5}(b) can account 
for the observed diffuse scattering pattern.

Figure \ref{Fig6} confronts the reciprocal space patterns of the 
HK0 and HK4 planes with the Fermi surface nesting construction. 
It can be appreciated that most of the non-TDS diffuse features can 
be associated to efficient nesting vectors (dark grey in the right 
panels). The parts of the squares and arcs observed in the diffuse 
scattering, and indicated by the dashed lines, can be found in 
the nesting constraction. Also the strong arcs inside the squares 
are fully reproduced. The spots with high intensity close to the 
$X$ points are indicated by circles. Not all features obtained from 
calculations can be observed in the diffuse scattering patter as 
the nesting construction does not take into account the modulation 
related to the x-ray structure factor of the spinel structure, which 
should attenuate/suppress a number of features compared to others. 
The real $Q$ dependence would take into account the structure factor 
arising from the average crystallographic positions and polaron-related 
displacements.

\begin{figure}[t!]
\centerline{\includegraphics[width = .48\textwidth]{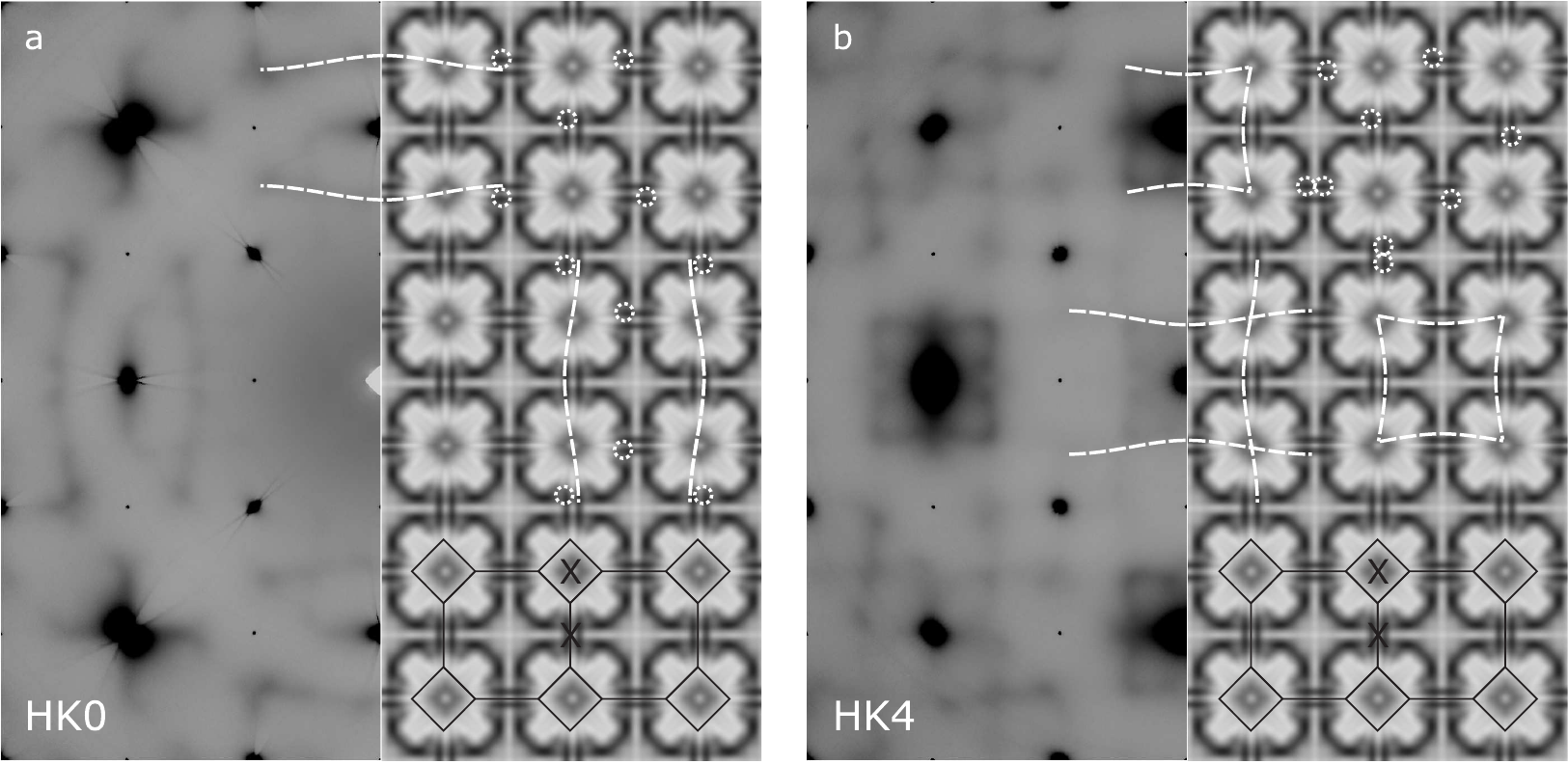}}
\caption{Nesting signatures in the reciprocal space of magnetite. 
Selected reciprocal space patterns of magnetite at $T_V+\delta$ 
(left subpanels) in the (a) HK0 plane and (b) HK4 plane are compared 
to the Fermi surface nesting construction (right subpanels). Dark 
features in the nesting construction correspond to more efficient 
nesting. Common features are underlined or encircled by dashed lines.
The contours of the Brillouin zone are superposed on the right 
subpanels, $X$ points are indicated. Note that the $X$ point is
avoided, both in the experiment and in the calculation.}
\label{Fig6}
\end{figure}

The absence of an efficient nesting vector directing to the $X$ point 
is reflected in the displacement of the maximum of diffuse scattering 
away from this point. The origin of this displacement is not clear 
but it points towards an incommensurate character of short-range 
fluctuations and is consistent with the recent inelastic x-ray 
scattering studies \cite{hoesch}. 

We emphasize that the agreement between the diffuse scattering data 
and the calculated Fermi surface nesting, shown in Fig. \ref{Fig6}, 
is remarkable. The matching between the observed data and calculated 
patterns appears to be very good indeed and this provides clear 
evidence that scattering within the minority spin $d$-band accounts 
for the diffuse scattering, and hence Fermi surface nesting may in 
part explain the nature of the Verwey transition.

\section{Summary}
\label{sec:summa}

In summary, our diffuse scattering study of magnetite allowed us 
linking the nature of short-range charge ordering above the Verwey 
temperature with the long-range structure of the low-$T$ phase. 
The short-range correlations generate a very rich pattern in large 
areas of reciprocal space with the highest intensities shifted from 
commensurate wavelengths. The complexes of trimerons are inherited 
from the low-$T$ structure with minor modifications. Their
characteristic correlation length can be in the order of $\sim 1.5$ nm 
just above $T_V$, and a local symmetry lower than cubic can be assumed. 

In this context the study of external stimuli on the diffuse scattering 
pattern (i.e., magnetic field) might be interesting, as it could reduce 
the symmetry of diffuse scattering without major changes in the average 
cubic structure. The coupling between charge fluctuations and lattice 
distortions (phonons) leads ultimately to the structural phase transition
with a complicated charge distribution. The underlying mechanism for the 
formation of nontrivial charge ordering can be related to nesting 
features of the Fermi surface and thus to the Fermi surface shape. 
We have previously demonstrated an example of successful use of Fermi 
surface reconstruction via 3D reciprocal space mapping associated 
with the electron-phonon coupling \cite{bosak}, and now we show that 
this approach can be extended well beyond the ''good'' metals. 

The most straightforward methods of Fermi surface measurements would 
fail in the case of magnetite --- for example, the de Haas-van Alphen 
and Shubnikov-De Haas effects, as well as positron annihilation cannot 
be used due to the required low temperatures at which Fe$_3$O$_4$ is 
insulating. The use of photoemission spectroscopy is far 
from being obvious due to the highly polar surface terminations that 
distort the electronic structure within the shallow probing depth, 
and potentially due to the surface instability under UHV conditions. 
Thus, providing constraints on the Fermi surface of magnetite from 
the diffuse scattering appears to be particularly attractive.

\acknowledgments

A.B. acknowledges Clemens Schulze-Briese for the access to the beamline 
X06SA at the SLS, and Daniele De Sanctis for the access to the beamline 
ID29 at the ESRF. 
P.P. and A.M.O. kindly acknowledge support by the Polish National 
Science Center (NCN) under Projects 
No. 2011/01/M/ST3/00738 and 
No. 2012/04/A/ST3/00331.

\end{document}